\documentclass{gGAF2e}
\usepackage{graphicx,natbib,bm,url,color}
\usepackage[colorlinks,allcolors=blue]{hyperref}
\graphicspath{{./figs/}{./png/}}

\newcommand{\brac}[1]{\langle #1 \rangle}
\newcommand{\EQ}{\begin{equation}}
\newcommand{\EN}{\end{equation}}
\newcommand{\EQA}{\begin{eqnarray}}
\newcommand{\ENA}{\end{eqnarray}}

\newcommand{\Sec}[1]{section~\ref{#1}}

\newcommand{\Fig}[1]{figure~\ref{#1}}

\newcommand{\Figs}[2]{figures~\ref{#1} and \ref{#2}}

\newcommand{\Tab}[1]{table~\ref{#1}}

\newcommand{\bra}[1]{\langle #1\rangle}

{}
{}
{}

{}
{}
{}
{}
{}
{}
{}
{}
{}
{}
{}
{}
{}
{}
{}
{}
{}
{}

{}
{}
{}

{}
{}
{}
{}

{}

{}
{}

\newcommand{\PC}{{\sc Pencil Code}~}
\newcommand{\gggg}{\mbox{\boldmath $g$} {}}

\newcommand{\II}{\mbox{\boldmath $I$} {}}

\newcommand{\qq}{\mbox{\boldmath $q$} {}}
\newcommand{\tqq}{\tilde{\mbox{\boldmath $q$}}{}}{}

\newcommand{\uu}{\mbox{\boldmath $u$} {}}

\newcommand{\BB}{\mbox{\boldmath $B$} {}}

\newcommand{\jj}{\mbox{\boldmath $j$} {}}
\newcommand{\JJ}{\mbox{\boldmath $J$} {}}

\newcommand{\AAA}{\mbox{\boldmath $A$} {}}

\newcommand{\KKK}{\mbox{\boldmath ${\cal K}$} {}}

\newcommand{\nab}{\mbox{\boldmath $\nabla$} {}}

\newcommand{\SSSS}{\mbox{\boldmath ${\sf S}$} {}}

\newcommand{\DD}{{\rm D} {}}

\newcommand{\dd}{{\rm d} {}}

\def\cs{c_{\rm s}}
\def\cp{c_{\rm P}}
\def\cv{c_{\rm V}}

\def\vA{v_{\rm A}}

\newcommand{\s}{\,{\rm s}}

\newcommand{\m}{\,{\rm m}}
\newcommand{\km}{\,{\rm km}}

\newcommand{\kms}{\,{\rm km/s}}

\newcommand{\gma}{\gamma_{\rm A}}

\definecolor{upforestgreen}{rgb}{0.0, 0.55, 0.13}

\begin{document}

\jvol{00} \jnum{00} \jyear{2018} 

\markboth{\rm {J.~WARNECKE AND S.~BINGERT}}{\rm {GEOPHYSICAL $\&$ ASTROPHYSICAL FLUID DYNAMICS}}


\title{Non-Fourier description of heat flux evolution in 3D MHD
  simulations of the solar corona}

\author{J\"orn Warnecke${\dag}$$^{\ast}$\thanks{$^\ast$Corresponding
    author. Email: warnecke@mps.mpg.de\vspace{6pt}} 
and Sven Bingert${\ddag}$\\\vspace{6pt}  
${\dag}$Max Planck Institute for Solar System Research,
Justus-von-Liebig-Weg 3, D-37077 G\"ottingen, Germany\\ 
${\ddag}$Gesellschaft f\"ur wissenschaftliche Datenverarbeitung mbH
  G\"ottingen, Am Fa\ss berg 11, D-37077 G\"ottingen\\\vspace{6pt}\received{\today}}

\maketitle

\begin{abstract}
The hot loop structures in the solar corona can be well modeled by three
dimensional magnetohydrodynamic simulations, where the  corona is heated by
field line braiding driven at the photosphere. To be able to reproduce the
emission comparable to observations, one has to use realistic values for the Spitzer
heat conductivity, which puts a large constraint on the time step of these
simulations and make them therefore computationally expensive.
Here, we present a non-Fourier description of the heat flux evolution, which allow
us to speed up the simulations significantly. Together with the
semi-relativistic Boris correction, we are able to limit the time step
constraint of the Alfv\'en speed and speed up the simulations even
further. We discuss the implementation of these two methods
to the \PC and present their implications on the time step, and the
temperature structures, the ohmic heating rate and the emission in
simulations of the solar corona.
Using a non-Fourier description of the
heat flux evolution together with the Boris correction, we can increase the time step of
the simulation significantly without moving far away from the reference
solution.
However, for values of the Alfv\'en speed limit of 3 000 $\kms$ and below, the
simulation moves away from the reference solution und produces much
higher temperatures and much structures with stronger emission.


\end{abstract}

\section{Introduction}

The solar corona can be described as a low  $\beta$ plasma at low densities and high
temperatures. With the presence of coronal magnetic fields, this leads
to plasma, where the magnetic pressure is higher than the
gas pressure.
Therefore, the plasma
motions are dominated by the magnetic field, and the plasma can organise
itself in accordance to the geometry of the magnetic field, e.g. closed
loop structures.
The hot plasma in the corona emits radiation in extreme UV and X-ray emission, making
it observable from space-based telescopes.
One of the major open questions concerning the solar corona is its
heating mechanism, i.e. why is the solar corona typically more than 100
times hotter than the photosphere.
One of the ideas explaining coronal heating is the field-line
braiding model by \cite{Par72,Par88}, in which magnetic energy is
released in form of nanoflares.
In this model, the magnetic footpoints of the loops are
irreversibly moved by the small-scale photospheric motions, get
braided in chromosphere and corona, where the reconnecting 
field lines release magnetic energy through ohmic heating and
contribute to the thermal energy budget.

Three-dimensional magnetohydrodynamic simulations modelling the
solar corona were able to show that with this nanoflare heating
mechanism the basic temperature structure and its dynamics can be
reproduced \citep[e.g.][]{GN02,GN05a,BP11}.
These models are able to describe energy transport to the corona
consistent with the nanoflare model \citep{BP13}.
These types of simulations are further used to synthesise coronal
emission comparable with actual observations of the corona.
From these synthesised emission, one finds that these models are able
to reproduce the average Doppler shifts to some extent \citep{PGN04,PGN06,HHDC10} and
the formation of coronal loops, when using a data driven model with an
observed photospheric magnetic field \citep{BBP13,BBP14,WP19}.
Furthermore, these models were used to show that the coronal magnetic field structure
is close to a potential field \citep{GN05a,BP11,BSB18}, and therefore
nearly force free.
However, the force-free approximation, broadly used to
obtain coronal magnetic field with field extrapolations \citep[for a
review, we refer to][]{Wi08}, turns out to
be not always valid \citep{PWCC15} and fails to describe complex
current structures in coronal loops above emerging active regions \citep{WCBP17}.
Recently, \cite{Rempel17} showed that the solar corona can be heated
by a small-scale dynamo operating in the near-surface region of the convection zone
braiding the magnetic footpoints in the photosphere.
Therefore, these types of models are able to reproduce the main
properties of the solar corona on the resolved scale
\citep[e.g.][]{P15}.
One of the most important ingredients is the vertical Poynting flux at the
bottom of the corona \citep[e.g.][]{GN96,BP11,BBP15}.

Currently there are only a limited number of codes available which are used
for this kind of simulations.
One of the most used codes to simulate the solar corona is the {\sc
  Bifrost} code \citep{GCHHLM2011}, which is based on
earlier work of  \cite{GN02,GN05a,GN05b} and the {\sc Stagger} code
\citep{GN96}. In these simulations, the near-surface convection is
self-consistently included and produces realistic photospheric velocities.
Furthermore, the {\sc Bifrost} code includes a realistic treatment of
the chromosphere using a non-local thermal equilibrium description.
Another code is the {\sc MuRAM} code \citep{VSSCEL05,Rempel14}
that has been recently extended to the upper atmosphere
\citep{Rempel17}. Also, there, the photospheric motions are driven by
near-surface convection.
Apart of these codes there are other codes used for realistic modelling of
the solar corona \citep[e.g.][]{Abbett07,MMLL05,MMLL08,HSMJMTG}.

In this paper, we present an extension to the coronal model of the {\sc
Pencil Code}\footnote{{\tt http://github.com/pencil-code}} that 
has been used successfully to describe the solar corona using either 
observed magnetograms and a velocity driver mimicking the photospheric
motions \citep{BP11,BP13,BBP13} or flux emergence simulations
\citep{CPBC14,CPBC15} as input at the lower boundary instead of
simulating the near-surface convection.
However, \cite{CP18} developed a 2D model, where the near-surface
convection is included with a realistic treatment of the solar corona.
Simplified two-layer simulations of the convection zone and the corona
of the Sun and stars using the \PC have been successfully used to investigate the
dynamo-corona interplay \citep{WB14,WKKB16}, to self-consistently drive current
helicity ejection into the corona \citep{WB10,WBM11,WBM12,WKMB13} and the formation of
sunspot-like flux concentrations \citep{WLBKR13,WLBKR16,LWBKR17}.

To be able to compare the simulations of the solar corona with
observations of emissivities, one needs to use a realistic value of the
Spitzer heat conductivity. However, this puts a major constraint on the
time step in these simulations. For simulations with a grid spacing of around
$200\,$km the time step due to the Spitzer heat conductivity is around
$1\,$ms.
However, this can be significant lower, if one does not limit the
diffusion speed by the speed of light.
If one wants to study the dynamics on smaller scales and being
able to reduce the fluid and magnetic diffusivities, one needs to use a
higher resolution. The smaller grid spacing leads to even lower values
of the time step. As the time step decreases quadratically with the grid
spacing, the simulations become unfeasible for very high resolutions.
To circumvent this,
\cite{CPBC14}, for example, used a sub-stepping scheme and
\cite{Rempel17} used a non-Fourier scheme, where the hyperbolic
equation for the heat transport is solved.
Similar approaches have also been used in the dynamo community to describe
the non-local evolution of the turbulent electromagnetic force
\citep{BKM04,HB09,RB12,BC18}.
We present here a non-Fourier description of the Spitzer heat flux, that
has been recently implemented to the {\sc Pencil Code}, see
\Sec{sec:heatflux}. We compare the outcome of the simulations obtained with and without the
non-Fourier scheme, see \Sec{sec:results}.
Furthermore, we also compare these simulations to those using
the semi-relativistic Boris correction \citep{Boris1970} to the Lorentz
force that has been also recently implemented to the {\sc Pencil Code}
\citep{CP18} to limit the time step constraint due to the Alfv\'en speed,
see \Sec{sec:boris}. 

\section{Setup}
\label{model}
The setup of the simulations is based on the model of \cite{BP11,BP13},
therefore a detailed description will not be repeated here.
We model a part of the solar corona in a Cartesian box ($x$,$y$,$z$) of
$100\times100\times60\,$ Mm$^3$ using a uniform grid.
The $z=0$ layer represents the solar
photosphere. We use $128\times128\times256$
grid points, corresponding to a resolution of $781\km$ in the
horizontal $234\km$ in the vertical direction.
We solve the compressible magnetohydrodynamic equations for the
density $\rho$, the velocity $\uu$, the magnetic vector potential
$\AAA$ and the temperature $T$.
\begin{equation}
{\DD\ln\rho\over\DD t} = -\nab{\bm \cdot}\uu,
\end{equation}
\begin{equation}
{\DD\uu\over\DD t} = -{\nab p\over\rho} +\gggg + {\JJ\times\BB\over\rho}
+ {1\over\rho}\nab{\bm \cdot}2\nu\rho\SSSS,
\end{equation}
\begin{equation}
{\DD\ln T\over\DD t} + \left(\gamma-1\right)\nab{\bm \cdot}\uu =
{1\over\cv\rho T} \left[\mu_0\eta\JJ^2 + 2\rho\nu\SSSS^2 -
  \nab{\bm \cdot}\qq + L \right]
\label{eq:temp}
\end{equation}
where we use a constant gravity $\gggg=(0,0,-g)$ with $g=274\,$m/s$^2$,
a rate of strain tensor
$\SSSS=1/2(u_{i,j}+u_{j,i})-1/3\delta_{ij}\nab{\bm \cdot}\uu$ and a constant
viscosity $\nu$ throughout the domain. Additionally we use a shock
viscosity to resolve shocks formed by high Mach number flows
(see \citealt{HBM04} and \citealt{GSFSM13} for details regarding its
implementation). 
The pressure $p=(k_{\rm B}/\mu\m_{\rm p})\rho T$ is given by the
equation of state of an ideal gas, where $k_{\rm B}$, $\mu$ and
$m_{\rm p}$ are the Boltzmann constant, the molecular weight and the
proton mass, respectively. The corresponding adiabatic index
$\gamma=\cp/\cv$ is 5/3 for a fully ionised gas, with the specific heats at constant pressure $\cp$
and constant volume $\cv$.
The heat flux $\qq$ is given by anisotropic Spitzer heat conduction
\begin{equation}
\qq=-K_0 \left({T\over [K]}\right)^{\!5/2} {\BB\!\BB\over \BB^2} \nab T
\equiv -\KKK\nab T,
\label{eq:Fouier_heatf}
\end{equation}
which only gives a contribution aligned with the magnetic field and
$K_0=2\times10^{-11}\,$W(mK)$^{-1}$ is the value derived by
\cite{Spitzer:1962} assuming a constant Coulomb logarithm. In general,
the Coulomb logarithm and therefore $K_0$ depends weakly on the coronal plasma density.
We limit the heat conductivity tensor such that the corresponding heat
diffusion speed $\dd x/(|\KKK|/\rho\cp)$ is 10\% of the speed of light
with $\dd x$ being the grid spacing.
For some of the runs we replaced this equation by the hyperbolic
equation of the non-Fourier heat flux, see \Sec{sec:heatflux}.
Additionally to the anisotropic Spitzer heat conduction, we apply an
isotropic numerical heat conduction, which
is proportional  to $|\nab\ln T|$ and a heat conduction with a
constant heat diffusivity $\chi = K/\cp\rho$.
These additions are used to describe the heat flux in the lower part of
the simulation, where the temperature is significantly lower and
therefore the Spitzer heat conductivity is significantly smaller than
in the corona. It also makes the simulation numerically more stable.

The radiative losses due to the optically thin part of the atmosphere
are described by $L=-n_{\rm e} n_{\rm H} Q(T)$, where $n_{\rm e}$ and
$n_{\rm H}$ are the electron and hydrogen particle densities. $Q(T)$
describes the radiative losses as a function of temperature following the
model of \cite{CCJA89}, for details see \cite{Bingert2009}.

To fulfil the exact solenoidality of the magnetic field
$\BB=\nab\times\AAA$ at all times, we solve for the induction equation
in terms of the vector potential $\AAA$.
\begin{equation}
{\upartial \AAA\over\upartial t} = \uu\times\BB +\eta\nabla^2\AAA,
\label{eq:aa}
\end{equation}
where we use the resistive gauge, i.e. arbitrary scalar field
  $\phi$, which divergence can be added to the induction
  equation is chosen to be $\phi=\eta\nab{\bm \cdot}\AAA$.
The currents are given by
$\JJ=\nab\times\BB$ and $\eta$ is the magnetic diffusivity.

\begin{figure}
\begin{center}
\includegraphics[width=0.6\textwidth]{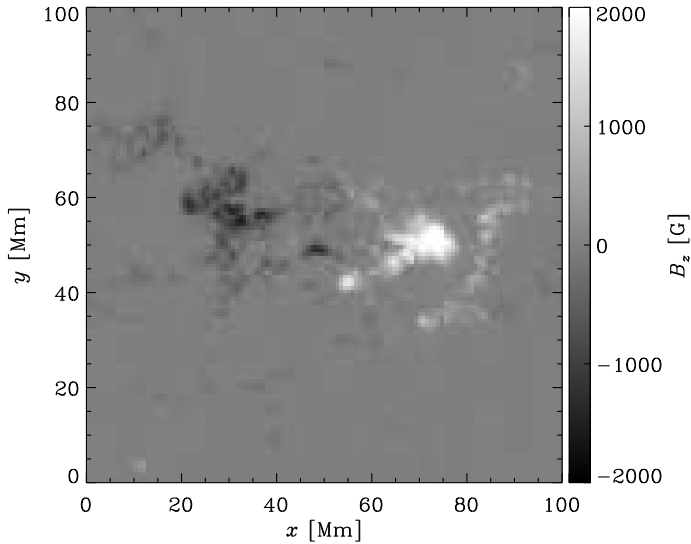}
\caption{Initial vertical magnetic field $B_z$ at the photospheric
  layer $z=0$ (colour online).
}\label{fig:bbz0}
\end{center}
\end{figure}

\subsection{Initial and boundary conditions}

At the lower boundary, we use for the vertical magnetic field the 
line-of-sight magnetic field from the active region
AR 11102, observed on the 30th of August with the Helioseismic and
Magnetic Imager \citep[HMI;][]{HMI} onboard of the Solar Dynamics Observatory
(SDO), see \Fig{fig:bbz0} for an illustration.
As an initial condition, we use a potential field extrapolation to
fill the whole box with magnetic fields.
For the temperature, we use an initial profile of a simplified
representation of the solar atmosphere, similar as in \cite{BP11}.
The density is calculated accordingly using hydrostatic
equilibrium. The velocities are initially set to zero.

The simulations are driven by a prescribed horizontal velocity field at
the lower boundary mimicking the pattern of surface convection.
As discussed in \cite{GN02,GN05a,Bingert2009} and \cite{BP11}, such a surface velocity
driver is able to reproduce the observed
photospheric velocity
spectrum in space and time.
To avoid the destruction of the magnetic field pattern caused by the
photospheric velocities, we apply the following to stabilise the
field: i) we lower the magnetic diffusivity in the two lowest grid
layers by a factor of 800 using cubic step function, ii) we
apply a quenching of velocities by a factor of 2, when magnetic pressure
is larger than the gas pressure and iii) we
interpolate between the current vertical magnetic field and the
initial one $B_z^{\rm int}$ at $z=0$ layer following
\begin{equation}
{\upartial B_z\over\upartial t}={1\over\tau_b} \left(B_z^{\rm int} -B_z\right),
\end{equation}
where $\tau_b=10\,$min is the relaxation time. 
The quenching of photospheric velocities mimics the suppression
  of convection in magnetised regions as observed on the solar
  surface  (see detailed discussion in \citealt{GN02,GN05a,Bingert2009} and
  \citealt{BP11}).
We apply a potential field boundary condition at the bottom and top
boundary of box for the magnetic field.
The temperature and density are kept fix at the bottom boundary.
The temperature is kept constant and the heat flux is set to zero at the top boundary
  allowing the temperature to vary in time. 
At the top boundary, we set all velocity components to zero to prevent mass
leaving or entering the simulation and to suppress all flows near the top
boundary. The density in the lower part is high enough to serve as a mass reservoir.
All quantities are periodic in horizontal directions.

For the viscosity we choose $\nu=10^{10}\m^2/\s$ similar to
  the Spitzer value for typical coronal temperatures and densities.
We set $\eta=2\times10^{10}\m^2/\s$ motivated by the numerical
stability of the simulations.
In the solar corona the magnetic Prandtl number $Pr_{\rm M}=\nu/\eta$
is around 10$^{10}$-10$^{12}$ and not 0.5 as in our simulations.

\subsection{Non-Fourier heat flux scheme}
\label{sec:heatflux}

To reduce the time step constraints due to the Spitzer heat conductivity,
we use a non-Fourier description and solve for the heat flux $\qq$
\begin{equation}
{\upartial \qq\over \upartial t} = -{1\over\tau_{\rm Spitzer}} \left(\qq +\KKK \nab
  T\right),
\label{eq:qq}
\end{equation}
where $\tau_{\rm Spitzer}$ is the heat flux relaxation time, i.e. e-folding time for $\qq$ to
approach $-\KKK \nab T$. $\KKK$ is the Spitzer
heat conductivity tensor, which has contributions only along the magnetic field.
This approach enables us to use a different time stepping constrain to
solve our equations.
Instead of using the time step of Spitzer heat conduction  $\dd t_{\rm Spitzer}=\dd
x^2/\gamma\chi_{\rm Spitzer}$ with $\chi_{\rm Spitzer}=|\KKK|/\rho\cp$, we
find two new time step constraints
\begin{equation}
\dd t_1=\dd x\sqrt{\left({\tau_{\rm
        Spitzer}\over\gamma \chi_{\rm Spitzer}}\right)} \equiv {\dd x\over c_{\rm
    Spitzer}}, \hskip 10mm \text{and} \hskip 10mm \dd t_2= \tau_{\rm Spitzer},
\label{eq:dt}
\end{equation}
where $\dd t_1$ comes from the wave propagation speed $c_{\rm Spitzer}$.
To see this more clearly, we can rewrite (\ref{eq:qq}) in one
  dimension $x$, with $q$ and $K$ being the one-dimensional
  counterparts of $\qq$ and $\KKK$ as
\begin{equation}
{\upartial q\over \upartial t} = -{1\over\tau_{\rm Spitzer}} \left(q +K {\upartial T\over \upartial x} \right)\,.
\end{equation}
Then together with a simplified one-dimensional
version of (\ref{eq:temp}), where we only consider the heat flux term
\begin{equation}
{\upartial T\over \upartial t} = -{1\over\cv\rho} {\upartial q\over \upartial x}\,,
\end{equation}
we can construct a wave equation for the temperature, namely
\begin{equation}
{\upartial^2 T\over \upartial t^2} = -{1\over\tau_{\rm Spitzer}}
  {\upartial T\over \upartial t} + {\gamma\chi_{\rm
        Spitzer}\over\tau_{\rm Spitzer}} {\upartial^2 T\over \upartial x^2},
\end{equation}
in which $c_{\rm Spitzer}$=$\sqrt{\gamma\chi_{\rm Spitzer}/\tau_{\rm Spitzer}}$
emerges as the propagation speed.
The two new time step constraints emerge from the pre-factors of the
terms on the right-hand side. 

By certain choices of $\tau_{\rm Spitzer}$, we can significantly increase the
time step. Furthermore, because $\dd t_1$ depends linear on the grid
spacing $\dd x$, instead of quadric as $\dd t_{\rm Spitzer}$, the
speed-up ratio $\dd t_1/\dd t_{\rm Spitzer}$ grows with higher
resolutions, which leads to a computational gain.
Both time step constraints are included in the CFL condition to
calculate the time step of the simulation.
$\dd t_1$ enters the time step calculation through the advective time $\dd t_{\rm advec}$ step using:
\begin{equation}
\dd t_{\rm advec}={\dd x\over u_{\rm advec}} \hskip 8mm  \text{with}\hskip 8mm  u_{\rm
  advec}=\max{\left(|\uu| + \sqrt{\cs^2+\vA^2} + c_{\rm
      Spitzer}\right)},
\label{eq:dtadvec}
\end{equation}
where $u_{\rm advec}$ is the advection speed and $\cs$ the sound
speed.

The major part of the heat flux is concentrated in the transition
region, where the temperature gradient is high. This can lead to
strong gradients in the heat flux $\qq$ itself. We, therefore,
normalise $\qq$ by the density $\rho$ to decrease the heat flux in the
lower part of the transition region compared to the upper part. 
The main motivation is to gain a better numerical
  stability and be able to resolve stronger gradients in $\qq$ better.
This results in a new set of equations, where
\begin{equation}
\tqq={\qq/\rho}.
\end{equation}
We basically solve now for the energy flux per unit particle instead of the
energy flux density.
\begin{equation}
{\upartial \tqq\over \upartial t}= {1\over\rho}{\upartial
  \qq\over \upartial t} - \tqq{\upartial\ln\rho\over \upartial t} =  -{1\over\tau_{\rm Spitzer}} \left(\tqq +{\KKK\over\rho} \nab
  T\right) + \tqq\left(\uu{\bm \cdot}\nab\ln\rho+\nab{\bm \cdot}\uu \right),
\label{eq:qqt}
\end{equation}
where we use the continuity equation to derive the last term.
The term in the energy equation changes correspondingly
\begin{equation}
{\upartial \ln T\over \upartial t} = -{1\over T\cv}\left(\nab{\bm \cdot}\tqq +
  \tqq{\bm \cdot}\nab\ln\rho\right) + \cdots\ .
\end{equation}
This formulation does not change the time step constraints shown in (\ref{eq:dt}).

Instead of choosing $\tau_{\rm Spitzer}$ as a constant value in time and space, we also implemented an
auto-adjustment, where $\tau_{\rm Spitzer}$ can vary in space and
time. This allows the simulation to be more flexible and to be able
to optimise the time step.
The main idea to choose a reasonable value for $\tau_{\rm Spitzer}$ is that we
set the time scale of the heat diffusion to be the smallest of all
relevant time scales in this problem, i.e. the
heat diffusion is the fastest process. The next bigger time scale is
typically the Alfv\'en crossing time $\dd t_{\rm vA} = \dd x /\vA$
with the Alfv\'en speed $\vA=B/\sqrt{\mu_0\rho}$. 
We want to keep the hierarchy of the time steps of each process in
place while lowering the time step as much as possible. 
So we choose the time step of the heat diffusion to be always a bit lower than the
  Alfv\'en time step, therefore the heat diffusion is still the fastest
  process, but slower as before. For a fixed ratio between the $\dd
  t_1$ and $\dd t_{\rm vA}$, we ``tie'' $\tau_{\rm Spitzer}$ to $\vA$ and we set
\begin{equation}
\dd t_1={\dd t_{\rm vA}\over\sqrt{2}}\hskip 5mm \rightarrow\hskip 5mm
c_{\rm Spitzer}=\sqrt{2}\vA \hskip 5mm\rightarrow \hskip 5mm \tau_{\rm
  Spitzer}={\gamma\chi_{\rm Spitzer}\over 2\vA^2}.
\label{eq:tau_auto1}
\end{equation}

On one hand, $\tau_{\rm Spitzer}$ would become very small in regions below the
corona, because there $\chi_{\rm Spitzer}$ has very low values due to
the low temperature and high density values. However, in these regions the
heat transport is mainly due to the isotropic heat transport. Low
values of $\tau_{\rm Spitzer}$ in these regions would cause a very small
time step, even though the Spitzer heat flux is not important for the
heat transport in these regions. Therefore, we choose the lower limit
to be the advective time step, which assures that $\tau_{\rm
  Spitzer}$ will not affect the time step in these regions.
On the other hand we want to avoid $\tau_{\rm Spitzer}$ becoming too
large and therefore the heat transport getting less efficient,
i.e. $\qq$ is still sufficiently close to $-\KKK \nab T$. So, we
choose $\tau_{\rm Spitzer}^{\rm max}=100\,$s as a limit for $\tau_{\rm
  Spitzer}$:
\begin{equation}
\min{\left(\dd t_{\rm vA}, {\dd x\over\sqrt{\cs^2+\uu^2}}\right)} \le
\tau_{\rm Spitzer} \le \tau_{\rm Spitzer}^{\rm max}.
\label{eq:tau_auto2}
\end{equation}

To use the non-Fourier heat flux description in the {\sc Pencil Code}, one has to
add \texttt{HEATFLUX=heatflux} to \texttt{src/Makefile.local} and set
the parameters in name list \texttt{heatflux\_run\_pars} in \texttt{run.in}.
The relaxation time $\tau_{\rm Spitzer}$ can be either chosen freely and
the inverse is set by using \texttt{tau\_inv\_spitzer} or one can
switch on the automatically adjustment by using
\texttt{ltau\_spitzer\_va=T}, then \texttt{tau\_inv\_spitzer} sets the
value of 1/$\tau_{\rm Spitzer}^{\rm max}$.

\subsection{Semi-relativistic Boris correction}
\label{sec:boris}

Above an active region the magnetic field strength can be high while
the density is low leading to Alfv\'en speeds comparable to the speed of
light \citep[e.g.][]{CF13,Rempel17}.
This causes two major issues. 
On one hand, the MHD approximation assuming non-relativistic phase speeds
is not valid anymore, i.e. we cannot neglect the displacement current.
On the other hand, the high values of the Alfv\'en speed reduce the
time step significantly.
To address these two issues we use a semi-relativistic correction of
the Lorentz force following the work of \cite{Boris1970} and
\cite{Gombosi02}, where we apply a semi-relativistic correction term to the
Lorentz force.
This has been used and successfully tested for the {\sc MuRAM} code in
\cite{Rempel17}. Here, we use the implementation discussed by
\cite{CP18}, who added this correction term to the {\sc Pencil
  Code}.
There, the Lorentz force transforms to
\begin{equation}
{\JJ\times\BB\over\rho} \hskip 4mm\rightarrow \hskip 4mm\gma^2 {\JJ\times\BB\over\rho} +
\left(1-\gma^2\right)\left(\II -
  \gma^2{\BB\BB\over\BB^2}\right)\left(\uu{\bm \cdot}\nab\uu + {\nab p\over\rho}
  -\gggg \right), 
\end{equation}
where $\gma^2=1/(1+\vA^2/c^2)$ is the relativistic correction factor. 
We note here that the correction term used here and in \cite{CP18} is
slightly different from the one used by \cite{Rempel17}, because
\cite{CP18} finds a more accurate way to approximate the inversion
of the enhanced inertia matrix. This leads to an additional $\gma^2$ in
front of $\BB\BB/\BB^2$.
If $\vA\ll c$ and $\gma^2\approx 1$, we retain the normal Lorentz force
expression. For  $\vA \le c$, the Lorentz force is reduced and the
inertia is reduced in the direction perpendicular to the magnetic field.
As the enhance inertia matrix \citep{Rempel17} is originally on the
right-hand side of the momentum equation, i.e. under the time derivative and it is just approximated by a correction
term on the left-hand side, the semi-relativistic Boris correction does not
change the stationary solution of the system and therefore does not lead
to further correction terms in the energy equation.
To switch on the Boris correction in {\sc Pencil Code} one sets the
flag \texttt{lboris\_correction=T} in the name list
\texttt{magnetic\_run\_pars}.

The Boris correction describes the modification of the Lorentz force
in the situation, where the Alfv\'en speed becomes comparable to the speed
of light.
In other words the speed of light is a natural Alfv\'en speed limiter
and the Boris correction describes the modification close to this
limiter.
We can artificially decrease the value of the limiter to a value of
our choice and the Boris correction takes care of the corresponding
modifications.
This can significantly reduce the value of the Alfv\'en speed in our
simulations and allow us to enhance the Alfv\'en time step.
Unlike in \cite{CP18}, we use the Boris correction indeed to increase
the Alfv\'en time step, similar to what has been done by
\cite{Rempel17}.
As shown by \cite{Gombosi02}, the propagation speed can be quite
complicated, we choose a similar time step modification as in \cite{Rempel17}
\begin{equation}
\dd t_{\rm vA} \hskip 4mm\rightarrow\hskip 4mm \dd t_{\rm vA}
\sqrt{1+\left({\vA^2\big/ c_{\rm A}^2}\right)^2},
\end{equation}
where $c_{\rm A}$ is the limiter. We choose for Set~B $c_{\rm A}=10\,000\kms$, which
corresponds to a time step of $\dd t_{\rm vA}\approx 20\,$ms for our simulations.
The limiter $c_{\rm A}$ can be set by using \texttt{va2max\_boris} in the name list
\texttt{magnetic\_run\_pars}.
The Boris correction can be used together with the automatic adjusted
relaxing time $\tau_{\rm Spitzer}$ in the non-Fourier heat flux
calculation: if one sets \texttt{va2max\_tau\_boris} in \texttt{heatflux\_run\_pars} to the same value as
\texttt{va2max\_boris} in \texttt{magnetic\_run\_pars}, then the code
modifies the Alfv\'en speed and the Alfv\'en time step used in
(\ref{eq:tau_auto1} and \ref{eq:tau_auto2}) accordingly.

\section{Results}
\label{sec:results}

We present here the results of three sets of runs, where we use
different values of the heat flux relaxation time $\tau_{\rm Spitzer}$
in combination with and without the Boris correction. In the first set,
containing only Run~R, we use the normal treatment of the Spitzer
heat flux without using the non-Fourier heat flux evolution and without the Boris correction.
In the second set, containing 4 runs (Set~H), we use the non-Fourier
heat flux
evolution with $\tau_{\rm Spitzer}$ between $10$ and $1000$ ms and the
automatically adjustment, see \Sec{sec:heatflux}.
In the third set, containing 7 runs (Set~B), we use the
semi-relativistic Boris correction with $c_{\rm A}=10\,000\kms$ and
the non-Fourier heat flux evolution with $\tau_{\rm Spitzer}=10-1000\,$ms and the automatically adjustment. We also use one run (Ba2)
with even lower Alfv\'en speed limit of $c_{\rm A}=3\,000\kms$.
An overview of the runs can be found in \Tab{runs}.

\begin{table}
\caption{Summary of the runs. $\tau_{\rm Spitzer}$ is the relaxation
  time for non-Fourier heat flux description, see \Sec{sec:heatflux}, $\tau_{\rm Spitzer}=\infty$
  stands for the use of standard Fourier heat flux, see (\ref{eq:Fouier_heatf}). $c_{\rm A}$ is the
Alfv\'en speed limit, used for the Boris correction, see
\Sec{sec:boris}; $c_{\rm A}=\infty$ stands for no Boris correction. $\dd t$
indicates the averaged time step, $\dd t_{\vA}$ the averaged Alfv\'en
time step and $\dd t_1$ and $\dd t_2$ the average time step due to
the heat flux evolution, see (\ref{eq:dt}). For Run~R, $\dd t_1=\dd
t_2=\dd t_{\rm Spitzer}$. All these quantities are determine
as an average in the quasi-stationary state. $t_{\rm cpu}$ is
  wall clock time per time step per mesh point. For the timing
  we use the SISU Cray XC40 supercomputing cluster at CSC. 
$\Delta T_{\rm cor}=(\brac{T}_{\rm runs}-\brac{T}_{R})/\brac{T}_{R}$ is the mean temperature 
deviation from the reference runs, taking as a horizontal and height ($z$=20-40 Mm) average.}
\vspace{1mm}
\small
\centering
\label{runs}
\begin{tabular}{lrrrrrrrr}

\hline\hline
Runs & $\tau_{\rm Spitzer}$ [ms] & $c_{\rm A}$ [km/s]& $\dd t$ [ms]&$\dd t_{\vA} $ [ms]&$\dd t_1$[ms]&$\dd t_2$ [ms]&$t_{\rm cpu}$ [$\mu$s]&$\Delta T_{\rm cor}$\\
\hline\hline
R & $\infty$ & $\infty$& 1.5& 2.7 & 1.7& 1.7& 4.2$\times 10^{-2}$& 0 \\   
\hline
H001 & 10 & $\infty$& 1.1& 2.8 & 1.2 & 9.0& 4.6$\times 10^{-2}$&-5\% \\    
H005 & 50 & $\infty$& 2.4& 3.0 & 4.4 & 45.0& 4.5$\times 10^{-2}$&-14\%\\  
H1 & 1000 & $\infty$& 4.5& 5.2 & 13.0 &900.0& 4.6$\times 10^{-2}$&-18\%  \\   
Ha & auto & $\infty$ & 2.0& 2.6 &  2.5 & 2.0& 4.6$\times 10^{-2}$&-3\%  \\   
\hline
B001 & 10 & 10 000 & 0.5& 21.2 & 0.5 & 9.0& 4.6$\times 10^{-2}$&14\%\\    
B002 & 20 & 10 000 & 2.8& 21.2 & 2.8 &18.0& 4.6$\times 10^{-2}$&-13\%\\    
B005 & 50 & 10 000 & 3.4& 21.2 & 3.7& 45.0& 4.5$\times 10^{-2}$&-7\%\\   
B01 & 100 & 10 000 & 4.4&21.2 &  4.7 & 90.0& 4.5$\times 10^{-2}$&4\%\\   
B03 & 300 & 10 000 & 5.4&21.2 &  6.1&270.0& 4.5$\times 10^{-2}$&-0.3\%\\   
B1 & 1000 & 10 000 & 8.0&21.2 & 10.0&900.0& 4.5$\times 10^{-2}$&-4\%\\   
Ba & auto & 10 000 & 15.4&21.2 &15.0&19.5& 4.7$\times 10^{-2}$&14\%\\  
Ba2 & auto & 3 000 & 47.6&70.6 &49.9 &64.9& 4.7$\times 10^{-2}$&18\%\\
\hline\hline
\label{tab1}
\end{tabular}
\end{table}

\begin{figure}
\begin{center}
\includegraphics[width=0.49\textwidth]{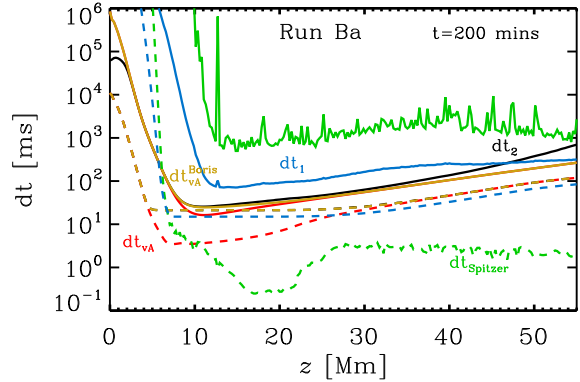}
\caption{Vertical distribution of time step
  constraints for Run~Ba at time $t=200$ mins. We plot the time step due to Spitzer
  heat conductivity $\dd t_{\rm Spitzer}$ (green), due to the heat flux
  $\dd t_1$ (black) and $\dd t_2$ (blue), due to the Alfv\'en speed
  $\dd t_{\rm vA}$ (red) and reduced Alfv\'en speed with the Boris
  correction $\dd t^{\rm Boris}_{\rm vA}$. The horizontal averaged values are shown with a
  solid line and the minimum values at each height with a dashed line (colour online).
}\label{fig:ts}
\end{center}
\end{figure}

\subsection{Time steps}

As a first step we look at the time steps of all the runs in \Tab{runs}.
In Run~R the averaged time step in the saturated stage is around
1.5 ms. This time step is constrained by the Spitzer time step $\dd
t_{\rm Spitzer}$, which is shown as $\dd t_1=\dd t_2$ in \Tab{runs}.
The Alfv\'en time step $\dd t_{\rm vA}$ is around twice as large.
In the Set~H, the code additionally solves the non-Fourier heat flux
equation that leads to an increased  time step. However, the time
step is actually limited by the low Alfv\'en time step and therefore
the time step cannot be increased by a large factor. 
In Set~H the largest speed-up factor is around 3.
For Run~H001, the value of $\tau_{\rm Spitzer}$ is low enough to have a
time step constraint of $\dd t_1$ instead of $\dd t_{\rm vA}$. However,
the runs reach a lower time step than in Run~R.
For values of the relaxing time $\tau_{\rm Spitzer}= 50-1000\ $ms
(Runs~H005 and H1), the time
step due to the heat flux is larger than the Alfv\'en time
step. This means that the physical process of heat redistribution is
even slower than the Alfv\'en speed.
This leads in Run~H1 to higher densities resulting in a lower Alfv\'en
speed and a higher $\dd t_{\rm vA}$ see discussion in \Sec{sec:emis}.
Furthermore, Run~H1 only runs stable, if we increase the shock viscosity to 10 times higher values than in the other runs. This
  will certainly lead to some additional differences independent of the
  direct influence of the non-Fourier heat flux description.
When applying the auto-adjustment of $\tau_{\rm Spitzer}$ (Run~Ha), the time steps
$\dd t_1$ and $\dd t_2$ are slightly smaller than $\dd t_{\rm vA}$ and
limits the time step.
There the speed up is less than a factor of 2, but the heat
distribution is the fastest process in the system.
Using the non-Fourier heat flux description leads
usually to higher peak temperatures, because the temperature diffusion
is less efficient.
For the calculation of $\dd t_1$ and $\dd t_2$, the code uses the CFL
pre-factors of $0.9$ for both time steps, this results in $\dd
t_2=0.9\, \tau_{\rm Spitzer}$. 
As $\dd t_1$ enters via (\ref{eq:dtadvec}), $\dd t$ is often lower than
$\dd t_1$ and $\dd t_{\rm vA}$ in our simulations.

To increase the time step further, we use the semi-relativistic
Boris correction in all runs of Set~B.
As shown in \Tab{runs}, $\dd t_{\rm vA}$ significantly increases to 21.2
ms for Runs~B001-Ba and to 70.6 ms for Ba2.
This leads to a much larger speed-up factor of 10 for Run~Ba and more than 30 for Run~Ba2.
For Run~B001 to Run~B1 with $\tau_{\rm Spitzer}$=
10-1000 ms, $\dd t_1$ is lower than $\dd t_{\rm vA}$ and the time step can
be significantly reduced, while the heat distribution is the fastest
process in the system.
For Run~B1, we achieve a speed up of  more than five, however we need to
use a comparable large value of $\tau_{\rm Spitzer}$, which as discussed
in \Sec{sec:emis} can lead to artefacts.
For Runs~Ba and Ba2, the auto-adjustment of  $\tau_{\rm Spitzer}$ takes
care that $\dd t_1<\dd t_{\rm vA}$.
As discussed below, Run~Ba shows a good agreement with Run~R, whereas
Run~Ba2 tends to produce higher temperatures in the corona.

To get a better understanding of the calculation of the time step, we plot in \Fig{fig:ts} various contributions to the time
steps for Run~Ba. 
Without the non-Fourier heat flux description and the Boris correction,
the time step is dominated by Alfv\'en time step $\dd t_{\rm vA}$ and
the Spitzer time step $\dd t_{\rm Spitzer}$.
The Boris correction reduces $\dd t_{\rm vA}$ to $\dd t^{\rm
  Boris}_{\rm vA}$ mostly in the regions between 5 and 30 Mm.
The auto-adjustment of $\tau_{\rm Spitzer}$ sets $\dd t_1$ to be always
slightly lower than $\dd t^{\rm Boris}_{\rm vA}$. 
Only below $z=5\,$Mm, $\dd t^{\rm Boris}_{\rm vA}$ is small, because
there the temperature diffusion is dominated by the other heat diffusion
mechanism described in \Sec{model}.
It is clearly visible that the $\dd t_1$ is significantly
higher than $\dd t_{\rm Spitzer}$ (green line) and $\dd t_{\rm vA}$
(red) without the Boris correction.
However, we note here that because of the non-Fourier heat flux
description we find higher peak temperatures in the simulation. This
results in a decrease of $\dd t_{\rm Spitzer}$ in comparison with runs
without the non-Fourier heat flux description. In Run~R, $\dd t_{\rm
  Spitzer}$ is around 1.6 ms, where in Run~Ba, it is around a factor
of eight lower. Such a factor can be explained by change in
temperature by a factor of 2.3. 

Using the non-Fourier heat flux evolution requires to solve (\ref{eq:qq})
or (\ref{eq:qqt}) meaning three additional equations. However, the
computational extra calculation time is around 10\%,  which is very
small compared to the gain in time step reduction. Using the
semi-relativistic Boris-Correction does not seem to increase the
computation time significantly. Only if we use the auto-adjustment of
$\tau_{\rm Spitzer}$ together with the Boris correction we find an additional
2-3\% increase in the computation time, as shown in the last row of \Tab{runs}.

\begin{figure}
\begin{center}
\includegraphics[width=\textwidth]{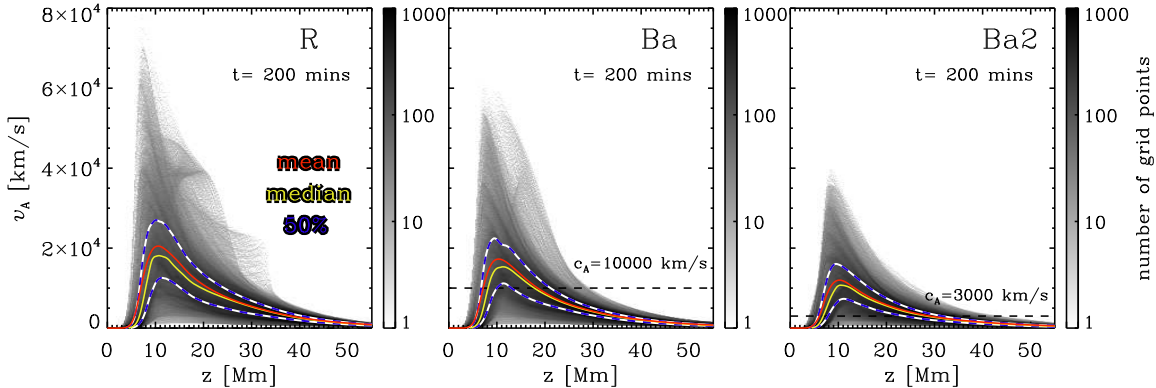}
\caption{2D histograms of the Alfv\'en speed over height $z$ for
  Runs~R, Ba, Ba2. We plot the mean value with red solid line and the 
  median  with a yellow solid. The dashed white-blue lines show the 25 and 75
  percentiles, i.e. half of the data points are in between these
  lines. The black dashed line
  indicate the Alfv\'en speed limit $c_{\rm A}$ for the Boris correction (colour online).
}\label{fig:alfven}
\end{center}
\end{figure}

\subsection{Alfv\'en velocity with Boris correction}

Next, we look at the influence of the semi-relativistic Boris
correction on the Alfv\'en velocity $\vA$.
In \Fig{fig:alfven}, we plot 2D histograms of $\vA$ for Runs~R, Ba,
Ba2. For Run~R, the maximum speed reaches $\vA=80\,000
\kms$ at the lower part of the corona, where the density has decreased
significantly with height, but the magnetic field is still strong.
The median (yellow line) has its maximum at the same location with a
value around $\vA=18\,000 \kms$.
In Run~Ba, we have applied the Boris correction with $c_{\rm
  A}=10\,000\kms$.
Even though, this value is lower than the averaged and mean value in the
region of $z=5$--$20$ Mm, the velocity distribution does not change
significantly in comparison to Run~R. As a main effect of the Boris
correction, the peak velocity at the top of the distribution is
reduced, therefore the distribution becomes more compact.
This can be also seen from the changes in the mean and
median velocity. While the maximum of the mean is reduced from above
$\vA=20\,000 \kms$ of Run~R to nearly $\vA=15\,000 \kms$, the
  median changes just slightly.
Also, the area between the 25 and 75 percentiles of the
  Alfv\'en velocity population moves only slightly towards lower
  values.
This make us confident that the Boris correction with $c_{\rm A}=10\,000\kms$ does only reduce the
peak velocities and not the overall velocity structure; most of the
points are unaffected by the correction.

For Run~Ba2, we reduce the Alfv\'en speed limit to $c_{\rm
  A}=3\,000\kms$. This makes the velocity distribution even more
compact.
The maximum values are significantly reduced to $\vA=35\,000 \kms$,
and the mean and median values are also lower than in Runs~R, Ba.
However, setting $c_{\rm A}=3\,000\kms$ does not mean that all the
velocities are lower than this value, it can be understood as a
significant reduction of the peak velocities and a transfer of the
velocity distribution to a much more compact form.

\begin{figure}
\begin{center}
\includegraphics[width=\textwidth]{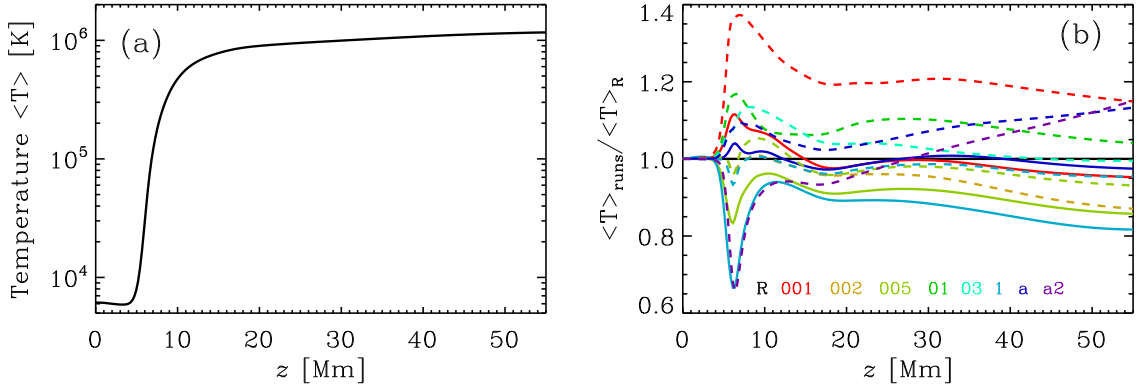}
\caption{(a) Averaged temperature $\bra{T}$ as a function of height $z$
  for Run~R. 
(b) Ratio of the averaged temperature profile of all runs and Run~R
$\brac{T}_{\rm runs}/\brac{T}_{R}$ as a function of height $z$. The temperatures are
  averaged horizontal as well as in time for the last quarter (1
  hours) of the simulation. The color of the lines indicates the run
  names in terms of $\tau_{\rm Spitzer}$, the solid lines are for
  runs of Sets~R and H, and dashed lines for Set~B (colour online).
}\label{fig:temp_prof}
\end{center}
\end{figure}

\subsection{Structure of temperature and ohmic heating}
\label{sec:temp}

\begin{figure}
\begin{center}
\includegraphics[width=\textwidth]{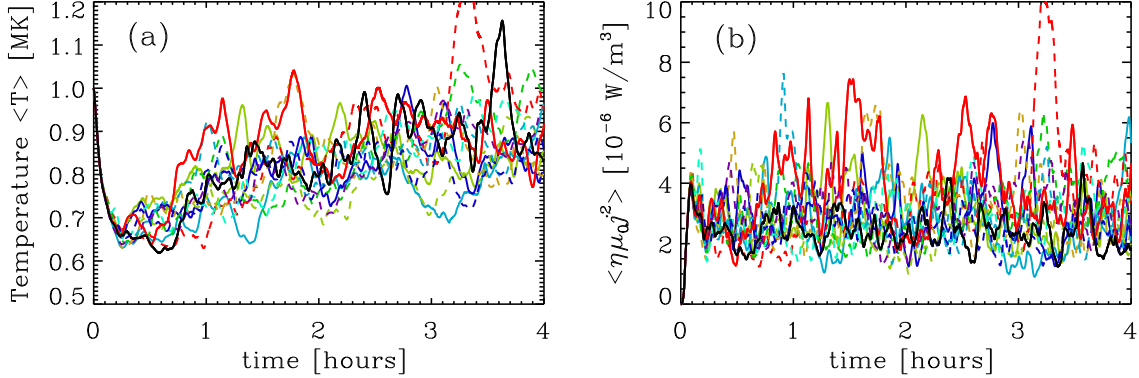}
\caption{Time evolution of the horizontal averaged temperature
  $\bra{T}$ (a) and of the horizontal averaged heating rate
  $\bra{\mu_0\eta\jj^2}$ (b) at $z=18-22\,$Mm. Color coding is the
  same as in \Fig{fig:temp_prof} (colour online).
}\label{fig:h_t_evo}
\end{center}
\end{figure}

Next, we look at the horizontal averaged temperature profile over
height. 
Even though the non-Fourier description of the heat flux can lead to higher
peak temperatures, the overall temperature structure should remain
roughly the same.
In \Fig{fig:temp_prof}, we plot the horizontal averaged temperature
profile over height for the reference Run~R in panel (a) and a
comparison with the other runs in panel (b).
The horizontal averaged temperature structure in Run~R shows a typical
behaviour of corona above an active region with medium magnetic field
strengths. 
The plasma above $z=10\,$ Mm is heated self-consistently to averaged
temperatures of around 1 million kelvin.
This temperature profile is very similar to results of earlier work with
the \PC \citep[e.g.][]{Bingert2009,BP11,BP13, BBP13} and other groups
\citep[e.g.][]{GN02,GN05a, GN05b, GCHHLM2011}.
When comparing with the temperature profiles of the other runs, we
find no large differences. For most of the runs the deviation is not
more than 10\%. For some runs the largest difference occurs
in the transition region, where the temperature has a large gradient.
Higher temperature values in this region simply mean a slightly lower transition
region and lower values mean a slightly higher transition region.
Nearly all runs develop a lower or similar transition region location
as in Run~R. Only Runs~H005, H1, Ba2 develop a higher transition region.
This can be explained either by sub-dominance of the heat flux time
step (Runs~H005, H1) or the too low limit for the Alfv\'en speed, see
discussion below.
Only in Runs~B001, Ba and Ba2 the plasma is heated to 20\% higher
temperature in the upper corona in comparison with Run~R.
For Run~B001, this high temperature only occur at the end of the
simulation, see \Fig{fig:h_t_evo}.
In these runs, the heat diffusion might be not efficient enough to
transport heat to lower layers.

When we look at the temperature evolution over time, as plotted in
\Fig{fig:h_t_evo}(a), we find that each run shows a large variation in
time even though we have averaged horizontally and over 18-20 Mm.
This can be explained by the non-linear behaviour of the
system. Because of this reason temporal variations occurring in the other runs
appear not at the same time for all runs.
The difference between the runs is comparable with the time variation
of each run.
Therefore, to be able to compare the runs, we should look at
the time averaged quantities as done throughout this work.

\begin{figure}
\begin{center}
\includegraphics[width=\textwidth]{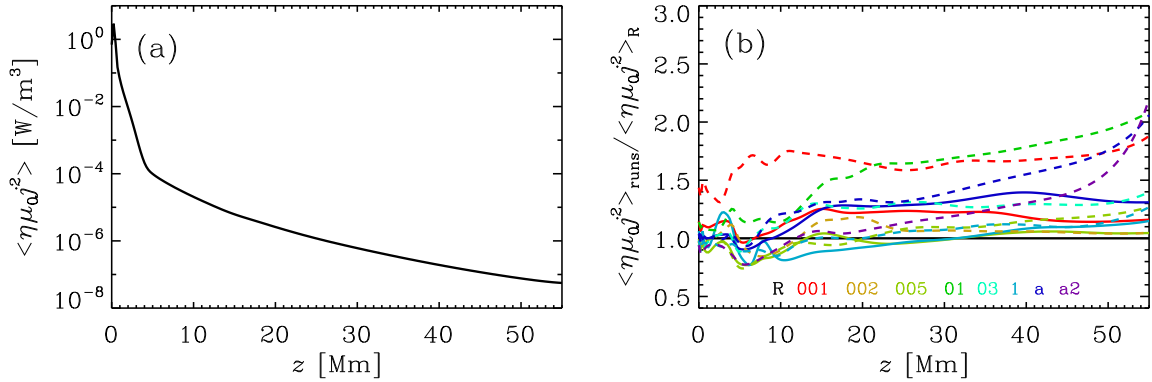}
\caption{(a) Averaged ohmic heating rate $\bra{\mu_0\eta\jj^2}$ as a function of height $z$
  for Run~R. 
(b) Ratio of the averaged ohmic heating rate of all runs and Run~R
$\brac{T}_{\rm runs}/\brac{T}_{R}$ as a function of height $z$. The ohmic heating rate are
  averaged horizontal as well as in time for the last quarter (1
  hours) of the simulation. The color coding the same as in \Figs{fig:temp_prof}{fig:h_t_evo} (colour online).
}\label{fig:heat_prof}
\end{center}
\end{figure}

Next, we look at the ohmic heating rate in all the runs.
The ohmic heating is the main process in this type of
simulations to heat the coronal plasma up to million K.
Also, here, we plot the horizontal averaged profile of Run~R in panel
(a) of \Fig{fig:heat_prof} and compare it with the other runs in \Fig{fig:heat_prof}(b).
The profile of the ohmic heating rate shows the typical behaviour of an
exponential decrease corresponding to two scale heights. Below the
corona the scale height is roughly 0.5 Mm, while in the corona the
scale height is around 5 Mm.
Also, this is consistent with earlier finding with this kind of
simulations by many groups \citep[e.g.][]{GN02,GN05a,
  GN05b,Bingert2009, GCHHLM2011, BP11,BP13, BBP13}.
By comparing with the other sets of runs, we find that these agree well with
Run~R.
Only Run~B001 shows a large heating rate in the lower corona, which
comes here also from the last part of the simulation. 
Runs~B01, Ba and Ba2 develop a higher heating rate in the
upper corona resulting in higher temperatures at this location ( see \Fig{fig:temp_prof}).
Small changes either in the scale height of the coronal heating or
in the location in the transition region can explain most of the
differences we find in the comparison with Run~R.
This explains also the temporal changes of the heating rate at
constant height, as shown in \Fig{fig:h_t_evo}(b).
The large variations in time of the heating rate can be attributed to
non-linear behaviour of the system. Even in Run~R, these variations are
large compared to the average. Small local changes in temperature and
density can also affect the heating rate. As the field is very close
to a potential field the currents are due to small perturbations from
the potential field. These perturbations can easily be affected by changes
in the plasma flow due to temperature and density fluctuations.
Furthermore, in such dynamical non-linear systems, changes
  for example in the time step can affect also the realisation of the
  velocity solution. Even when solutions are the same on a
  statistical level, this can cause variations in the ohmic heating.
For these types of models, large variations in time of the ohmic heating
rate are a common feature \citep[e.g.][]{BP11,BP13} as small changes in local scale
height will lead to a large change in the heating rate.
Overall, the vertical horizontally averaged temperature and heating
structure of all runs agree well with Run~R. 

\subsection{Emission signatures}
\label{sec:emis}

\begin{figure}
\begin{center}
\includegraphics[width=1.05\textwidth]{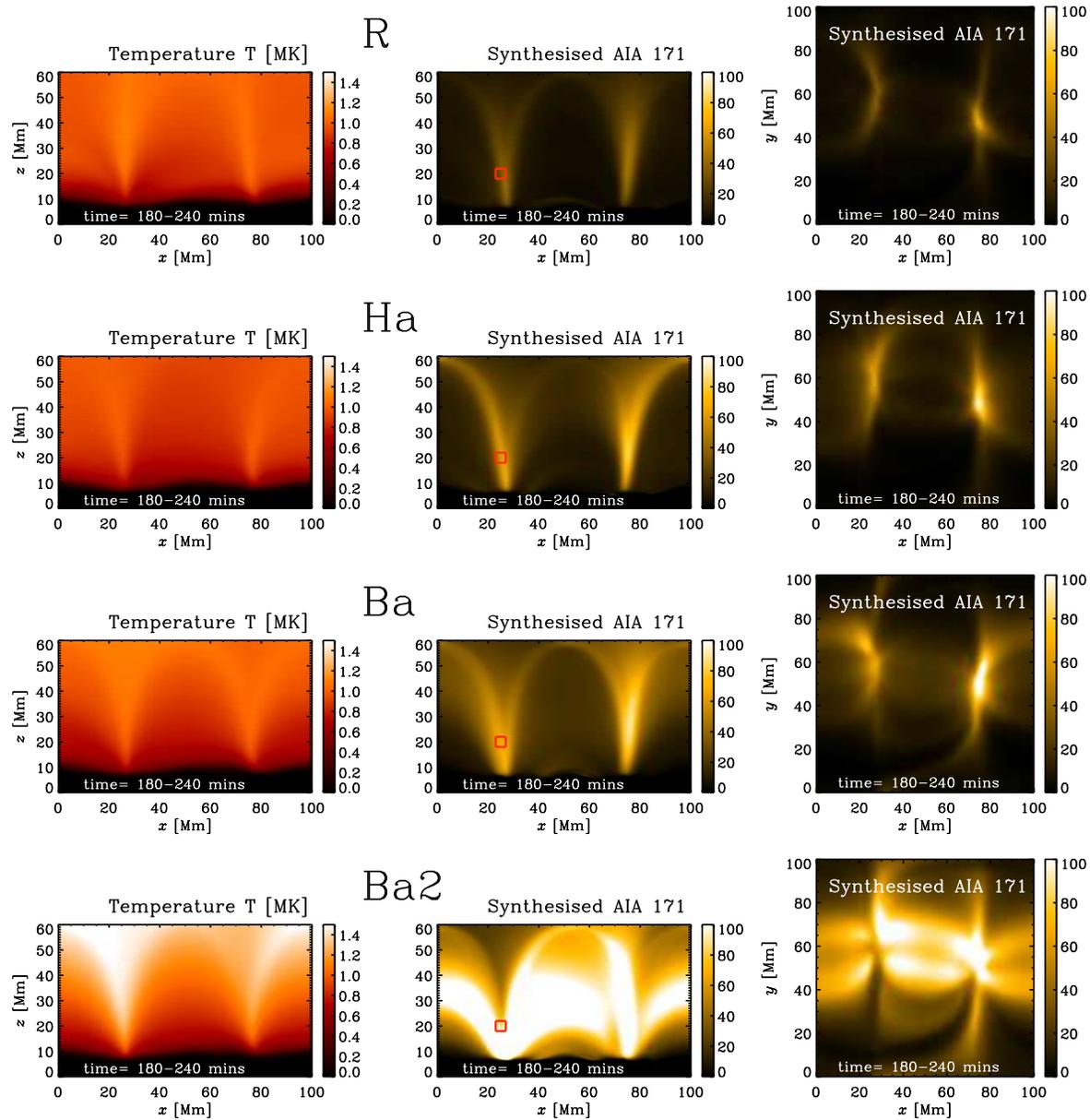}
\caption{Temperature and emission structure for Runs~R, Ha, Ba,
  Ba2. We show the temperature averaged over the $y$ direction
    and in time $t=180-240$ mins (left panel) together with the synthesised
  emission comparable to the AIA 171 channel, representing emission at
  around 1 MK, integrated in the $y$
  direction (side view, middle panel) and in $z$ direction (top view,
  right panel). The emission values represent the count rate of the
  AIA instrument and has been averaged in time $t=180-240$ mins.
The red square indicate the region which is used to calculate the
temporal evolution in \Fig{fig:emis_ts} (colour online) (colour online).
}\label{fig:emis1}
\end{center}
\end{figure}

\begin{figure}
\begin{center}
\includegraphics[width=0.49\textwidth]{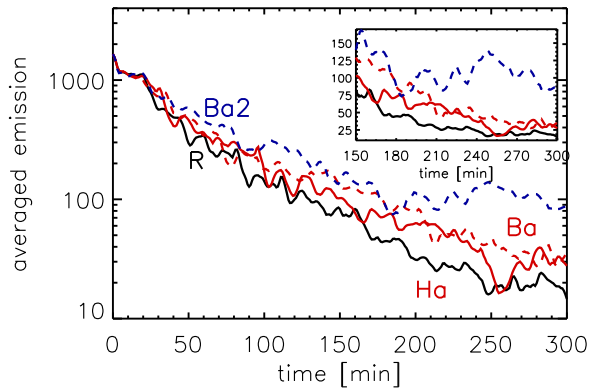}
\caption{Time evolution of emission averaged over a small region for
  Run~R, Ha, Ba and Ba2. We plot the emission of the AIA 171
  channel in the $y$ direction averaged over a small region ($x=23-27$ Mm,
  $z=18-22$ Mm) as indicated by red boxes in \Fig{fig:emis1} middle
  panels. The inlay shows the time evolution of the averaged emission
  from 150 to 300 mins on a linear scale instead of logarithmic. The
  color and style of the lines are the same as in \Fig{fig:temp_prof} (colour online).
}\label{fig:emis_ts}
\end{center}
\end{figure}

To further test how well the non-Fourier description of the heat flux
reproduce the Fourier description, we synthesise coronal emissivities
corresponding to the 171 \r{A} channel of Atmospheric
Imaging Assembly \citep[AIA;][]{AIA:2012} on board of SDO.
We choose this AIA channel because it can be potentially compared with
observations and represents well the plasma structure of around 1
million K by convolving the temperature and density structures.
This can work as a good test, weather or not coronal emission structures are
affected by the choice of heat flux description.
For this we calculate the emission following optical thin radiation approximation,
\begin{equation}
\epsilon=n_e^2 G(T),
\label{eq:emis}
\end{equation}
where $G(T)$ is the response function of the particular filter, we want
to synthesise. Because we compare our simulations among each
other and not to observation, we simplify $G(T)$ using a gaussian
distribution around a mean temperature $\log_{10} T_0$,
\begin{equation}
G(T) \propto \exp{\left[- \left({\log_{10} T - \log_{10}
        T_0\over\Delta\log_{10} T_0}\right)^2\right]},
\label{eq:emis2}
\end{equation}
where $\Delta\log_{10} T_0$ is the temperature width used to mimic the temperature
response function.
We use $\log_{10} T_0=6\, \log_{10}\, $K and $\Delta\log_{10} T_0=0.2\,
\log_{10}\, $K for
synthesising the emission of the AIA 171 \r{A} channel.
To calculate the emission emitted from a certain direction, we perform
  an integration along this direction. For the discussion below, we apply
  an integration along the $y$ and $z$ directions, respectively.

In \Fig{fig:emis1}, we plot the temperature as a side view ($xz$) averaged
over $y$ and in time (180-240 min) together with the
synthesised emission integrated over the $y$ and $z$
  directions also averaged in time (180-240 min) representing the AIA
171 channel for Runs~R, Ha, Ba, Ba2.
For these runs, we expect a good agreement with the reference run R,
because the value of $\tau_{\rm Spitzer}$ is regulated
automatically and therefore the time step is controlled by the heat flux,
i.e. $\dd t_1$.
We find agreement between the Runs~R, Ha  and Ba, but we find
slightly stronger emission structures in Run~Ba and slightly hotter
temperatures in Run~Ha. To illustrate the variation in time we show in \Fig{fig:emis_ts} the
time evolution of the emission in a small region of the simulation box.
We find a good agreement between Runs~R and Ha with variation in time
which are comparable with their difference.
Run~Ba takes a bit longer to saturate, but at around 220 min it also settles to values similar to
Runs~R and Ha. Run~Ba2 seems to saturate to a much higher emission
level than the other runs, which is already seen in \Fig{fig:emis1}.

For Run~Ba2, as pointed out in \Sec{sec:temp} and shown in first
column of \Fig{fig:emis1}, the corona is heated to
higher temperatures, i.e. the heat transport is less efficient.
We find larger temperatures mostly at the top of the corona inside the
loop structures. This leads also to higher emission in the AIA 171
channel than in the Run~R.
This might be an artefact from the low limit of the Alfv\'en speed through
the Boris correction in this run.
Even though the Alfv\'en speed limiter does not effect the
  heat flux directly, it increases the heat flux time step and makes
  the heat transport less efficient.

\begin{figure}
\begin{center}
\includegraphics[width=1.05\textwidth]{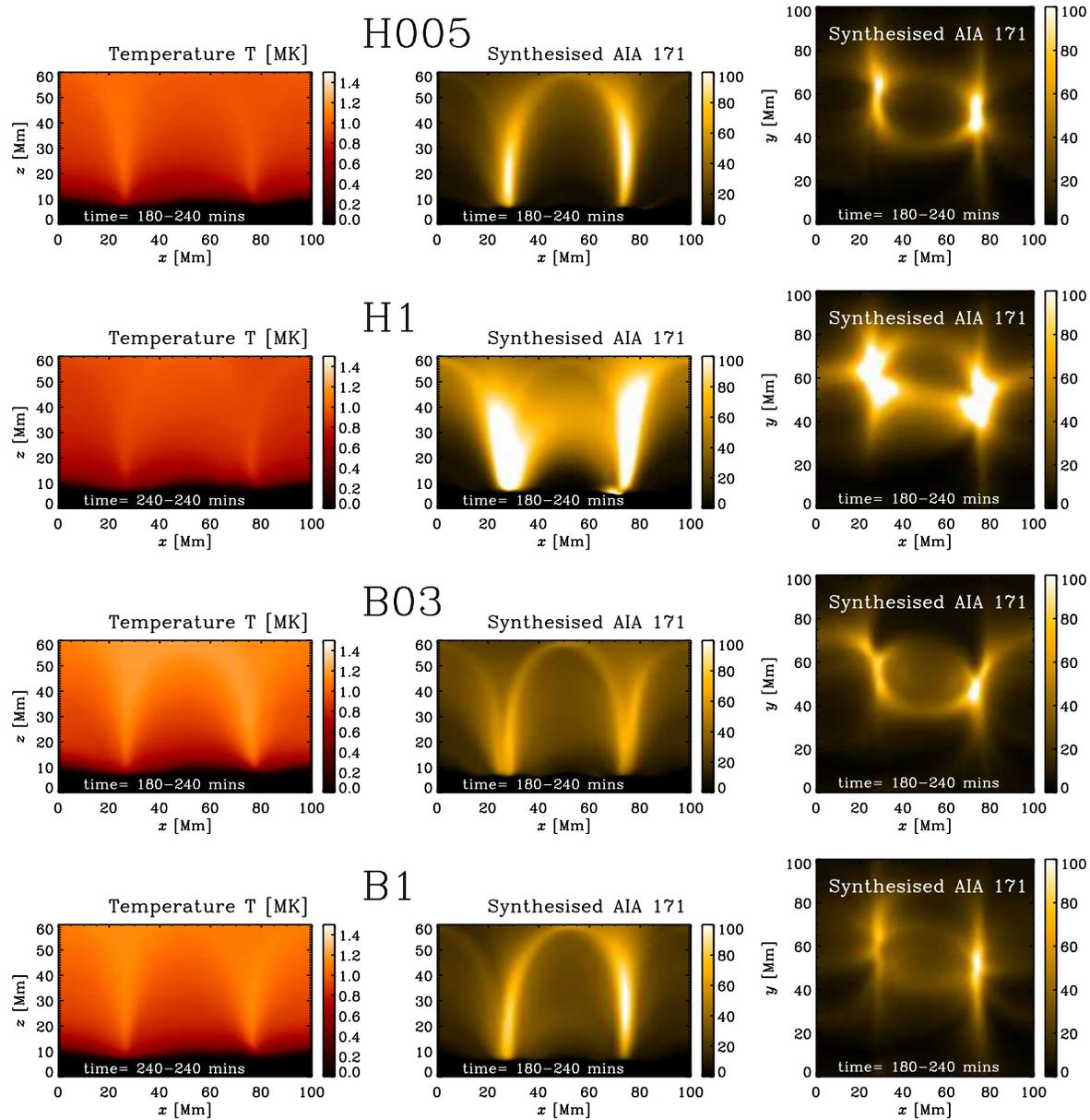}
\caption{Temperature and emission structure for Runs~H005, H1, B03,
  B1. The emission values represent the count rate of the
  AIA instrument. We show the temperature averaged over the $y$ direction
    and in time ($180-240$ mins, left panel) together with the synthesised
  emission comparable to the AIA 171 channel, representing emission at
  around 1 MK, integrated in the $y$
  direction (side view, middle panel) and in $z$ direction (top view,
  right panel). The emission values represent the count rate of the
  AIA instrument and has been averaged in time ($180-240$ mins) (colour online). 
}\label{fig:emis2}
\end{center}
\end{figure}

In \Fig{fig:emis2}, we show a few other runs, which are either dominated
by the Alfv\'en time step (Runs~H005 and H1) or use a constant value
of $\tau_{\rm Spitzer}$ (Runs~B03 and B1).
Run~H005 shows a similar emission structure than the Runs~R, Ha, Ba,
however the emission is slightly larger in the legs of the loop. Because
also here the temperatures are not significantly higher, the
difference is due to the slightly higher density in these regions.
For Run~H1, $\tau_{\rm Spitzer}$ is large and the time
step is controlled by the Alfv\'en speed instead of the of heat flux. This leads to larger temperatures and
therefore higher emission. However, also here the density in the corona
loops is larger than in Run~R, leading not only to higher emission,
but also to a larger Alfv\'en time step, see \Tab{tab1}.
Furthermore, the high shock viscosity needed to keep the run stable
will also have an influence on the solution.
In contrast, the time steps in Runs~B03 and B1 are controlled by the time step of
the heat flux ($\dd t_1$). There, as expected, we find similar emission
loop structures as in Run~R, Ha and Ba. They are slightly larger in
Run~B1 than in Run~B03.
This means that simulations using either the automatic adjustment or a constant
value of $\tau_{\rm Spitzer}$ reproduce the emission structure of
Run~R well, as long as the time step is still controlled by the heat flux
time step  $\dd t_1$, however the emission tends to be slightly larger.
However, for too low values of the Alfv\'en limiter
($c_{\rm A}=3000\kms$) the
emission and temperature become much higher than in Run~R.
We note here that the AIA 171 channel is relatively broad
  filter around the mean temperature and therefore hide some of the
  differences between the runs. A more narrow filters for example used
  on Hinode/EIS might reveal larger differences.

\section{Discussion and conclusion}

In this work we present the new implementation of a non-Fourier
description of the heat flux to the {\sc Pencil Code}. We discuss the advantages and the
limitations using the example of 3D MHD simulations of the solar corona.
The implementation of the auto-adjustment of $\tau_{\rm Spitzer}$ is
slightly different from the implementation used in \cite{Rempel17} in
the sense that we ensure the heat flux time step to be always by a square root
of two smaller than the  Alfv\'en time step,
whereas in \cite{Rempel17} there is not such a factor.
Even though a detailed comparison was not conducted here, we see
indications that our choice leads to a better stability of the
simulations.
We find that using the non-Fourier description of the heat flux alone
allows for a small speed up, because in our case the time constraint of the
Alfv\'en speed is large. For simulations with a lower magnetic
field strength, we would expect a larger speed up.
If we choose a constant $\tau_{\rm Spitzer}$, so that 
the heat flux time step is four times higher than the Alfv\'en time step,
the temperatures and the emission are significantly larger than in the
other runs. This seems to be an artefact of this choice of $\tau_{\rm Spitzer}$.

We further test the implementation of the semi-relativistic Boris
correction \citep{Boris1970} as a limiter for the Alfv\'en speed.
The implementation to the \PC is slightly different from the one used
by \cite{Rempel17} and \cite{Gombosi02}, see \cite{CP18} for details.
The Boris correction does not quench the Alfv\'en speed at all locations 
to the limit chosen, it actually reduces the peak velocities, which
are not very abundant. Therefore, this correction makes the velocity
distribution much more compact. The lower the limit, the more compact is the
velocity distribution.
Using the Boris correction allows for a significant speed up of around 10.
For higher speed up, i.e., lower limit for Alfv\'en speed, the simulation
develops higher temperatures and emission signatures than the reference
run. The auto-adjustment together with Boris correction works very
well to reproduce the temperatures and emission structures of the
reference run with a speed up of around 10 (Run~Ba).
These results convince us that we can use the non-Fourier heat flux
description together with the Boris correction to acquire a significant speed up
of the simulation without losing a correct representation of the physical
processes within the solar corona in a statistical sense.
We find some differences between the solution with and without non-Fourier heat flux
description and the Boris correction. However, we are not interested
if the non-Fourier heat flux description is identical to Fourier heat
flux description in every time step at every specific
location. Instead, we are interested if the non-Fourier heat flux
description reproduced the Fourier heat flux description on a
statistical level. On the statistical level we find a very good agreement.

In the future, we are planning to use these implementations to perform
large-scale active region simulations similar as done by \cite{BBP13,BBP14},
which can be then run for a much longer time and allowing the study of
hot core loop formations.
A first attempt is already published \citep{WP19}.
Furthermore, this implementation allows us to perform parameter
studies to investigate the coronal response to different types of
active regions on the Sun and also on other stars.
Finally, through these improvements, we get closer to the possibility
to simulate a more realistic convection-zone-corona model as started
in \cite{WKMB12,WKMB13,WKKB16}.

\section*{Acknowledgements}
We thanks the anonymous referees for the useful suggestions, we also
thank Piyali Chatterjee for the implementation of the Boris correction
to the \PC and detailed discussions about it. We furthermore thank Hardi Peter for discussion about the
non-Fourier heat flux description and comments to the manuscript.
The simulations have been carried out on supercomputers at
GWDG, on the Max Planck supercomputer at RZG in Garching, in the facilities hosted by the CSC---IT
Center for Science in Espoo, Finland, which are financed by the
Finnish ministry of education. 
J. W.\ acknowledges funding by the Max-Planck/Princeton Center for
Plasma Physics.

\markboth{\rm {J.~WARNECKE AND S.~BINGERT}}{\rm {GEOPHYSICAL $\&$ ASTROPHYSICAL FLUID DYNAMICS}}
\bibliographystyle{gGAF}
\markboth{\rm {J.~WARNECKE AND S.~BINGERT}}{\rm {GEOPHYSICAL $\&$ ASTROPHYSICAL FLUID DYNAMICS}}
\bibliography{paper}
\markboth{\rm {J.~WARNECKE AND S.~BINGERT}}{\rm {GEOPHYSICAL $\&$ ASTROPHYSICAL FLUID DYNAMICS}}

\end{document}